\documentclass[a4paper,11pt]{article}

\usepackage{amsmath}
\usepackage{amsfonts}
\usepackage{amssymb}
\usepackage{amsthm}
\usepackage{slashed}
\usepackage{braket}
\usepackage{mathrsfs}
\usepackage[margin=1in]{geometry}
\usepackage{bm}
\usepackage{siunitx}
\usepackage{hyperref}
\usepackage{wrapfig}
\usepackage{mathtools}
\usepackage{xcolor}
\usepackage{authblk}
\usepackage{algorithm}
\usepackage[noend]{algpseudocode}

\newtheorem{theorem}{Theorem}

\newtheorem{proposition}[theorem]{Proposition}

\newtheorem{definition}[theorem]{Definition}

\usepackage[autocite = superscript]{biblatex}
\AtBeginDocument{

\DeclareSymbolFont{bbold}{U}{bbold}{m}{n}
\DeclareSymbolFontAlphabet{\mathbbold}{bbold}

} 

\hfuzz=\maxdimen \tolerance=10000 \hbadness=10000

\title{Reduction Rules and ILP Are All You Need: Minimal Directed Feedback Vertex Set}
\author[1]{Alex Meiburg}

\affil[1]{Department of Physics, University of California, Santa Barbara, CA 93106-5070, USA}
\begin{document}
\maketitle

\begin{abstract}
   This note describes the development of an exact solver for Minimal Directed Feedback Vertex Set as part of the PACE 2022 competition. The solver is powered largely by aggressively trying to reduce the DFVS problem to a Minimal Cover problem, and applying reduction rules adapted from Vertex Cover literature. The resulting problem is solved as an Integer Linear Program (ILP) using SCIP. The resulting solver performed the second-best in the competition, although a bug at submission time disqualified it. As an additional note, we describe a new vertex cover reduction generalizing the Desk reduction rule.
\end{abstract}

\section{Introduction}
The 2022 Parameterized Algorithms and Computational Experiments challenge\autocite{pace2022} was to build an (exact or heuristic) solver for Directed Feedback Vertex Set: given a directed graph $G = (E,V)$, finding the smallest set of vertices whose removal leaves an acyclic graph. This problem has algorithms better than the naive $O(2^n)$ approach\autocite{Razgon07} and is in fact fixed-parameter tractable\autocite{Chitnis15,Lokh16}. Curious to see how effective (or not) Integer Linear Programming would be at attacking the problem, we built a solver\footnote{GitHub Link: \url{https://github.com/Timeroot/DVFS_PACE2022/}} based just on applying simple reduction rules and then an efficient MILP formulation. The solver was recorded to score second place among all submissions\autocite{paceResults}, although unfortunately was disqualified from prizes due to a bug at the time of submission. The purpose of this note is to describe the approaches used and the practical questions faced in design of this solver.

\section{Approach}
The central idea was that, as MILP solvers are very mature and flexible as a generic branch-and-bound framework, we should build our solver within this framework, and offload as much work to their intelligent heuristics as possible. Designing our own branch-and-bound framework, with all of the attached questions of efficient memory management, heuristic, branch prioritization, etc. can quickly outpace all the other development time. For the MILP solver itself, we used SCIP\autocite{Scip8} (at its latest version, 8), as SCIP is design with a plugin-first architecture that would allow custom constraint enforcement and heuristics. The simplest MILP formulation of DFVS is

\begin{align*}
    \textrm{Minimize:} \quad & \sum_i v_i\\
    \textrm{Subject to:} \quad & \forall_{\textrm{Directed Cycle}\, C \in G}\,\, \sum_{v_i \in C} v_i \, \ge 1\\
    & v_i \in \{0,1\}
\end{align*}

where the binary variable $v_i$ is associated to each vertex, and the constraints require at least one vertex selected in each directed cycle. However, this representation can be exponentially large, as there can be exponentially many directed cycles in the graph. This can be addressed with a lazy constraint handler, which checks at various points during the MILP solving for violated constraints, and marks nodes as infeasible or adds separating cuts as necessary. Alternate formulations could include constraints base on min-cut/max-flow to limit self-flow in the graph, or constraints imposing a topological order on the remaining graph. However, these create a quadratic increase in number of variables and have their own performance issues.

In practice, however, the above formulation may be quite manageable in size. This stems from the fact that only {\em minimal} (i.e. chordless) directed cycles need to be included in the MILP, and that many DFVS problems of practical interest are quite sparse. Additional reduction rules may allow the graph $G$ to be shrunk or sparsified to improve this further. The high-level architecture of our solver is then:

\begin{enumerate}
    \item Apply reduction rules to the original graph $G$ to shrink it, sparsify it, or split it into smaller graphs or subproblems.
    \item Try to identify the complete set of minimal cycles in the graph. As some minimal cycles are found and added to the running list, the graph may be further shrunk. This transforms the problem into a collection of "DFVS constraints" for the remaining graph(s), and cover constraints for the found cycles.
    \item Apply reduction rules to the cycle cover constraints. For size-2 cycles, this locally functions as a vertex cover problem, which has particularly well-studied reduction rules.\autocite{Fellows2018,Xiao13} For larger cycles or constraints that intersect graphs, more care is needed.
    \item The resulting problem is handed to the MILP solver. If there are no DFVS constraints, the MILP formulation is complete, and the result from the MILP solver is all we need. If there are DFVS constraints remaining, then we implement a lazy constraint handler to enforce them.
\end{enumerate}

To our surprise, our solver actually terminates at step 3 on a 36 of the 100 test instances, because the reduction rules from steps 1-3 are powerful enough to completely solve the problem. On an additional 46 test instances, all graphs were fully reduced to a small set of minimal cycles, removing the need for a lazy constraint handler. Being able to omit the constraint handler seems to significantly improve the performance of the MILP solver, as more aggressive presolving (essentially, reduction rules built into the solver) can be applied without concern for unknown lazy constraints.

The next section describe the rules used in steps 1-3, and Section 4 describes the MILP solver interaction.

\section{Reduction Rules}
The majority of the code written focuses on reduction rules. The first set of operations are performed on the input graph itself, transforming into smaller graph(s) before the cycle cover reduction.
\subsection{Graph Rules}
After a few experiments, only a few kernelizing rules were used on the DFVS graph. Some others were considered, but would have incurred significant development time (and bug-prone code complexity), and could be more easily applied to the reduced cycle cover instead. The only rules used were:
\begin{enumerate}
\item Split the graph into Strongly Connected Components (SCCs) and solve each separately as a subproblem.
\item Vertices with in-degree or out-dedgree one can be contracted with their unique neighbor.
\item Vertices with in- or out-degree zero can be removed.
\item Vertices with a self-loop must be included in the solution and can then be removed.
\end{enumerate}
All these rules are well known, for instance, Rule 3 above is Reduction Rule 1 in \cite{Bergougnoux2021}, and Rule 2 is their Reduction Rule 2. Rules 2-4 are efficiently implemented with a queue of "vertices to be checked" that ensures linear time kernelization, despite the fact that each can trigger new applications of the other. After this, Rule 1 is applied with Tarjan's algorithm\autocite{TarjanSCC}, and then another linear pass of Rules 2-4. Although this could lead to new SCC splitting, this was never observed in practice, so subsequent re-checking of SCC splits was omitted in the final solver.

Other rules considered for inclusion were:
\begin{enumerate}
    \item Twins - If two vertices have identical in-neighbors and out-neighbors, then either both must be included in the solution, or neither. This could simplify the graph, but would effectively turn into a significantly more complicated {\em weighted} cover problem.
    \item Vertex Domination - If there are vertices $u$ and $v$ such that all descendents of $u$ in $G\setminus\{v\}$ form an acyclic graph, then $u$ can be removed from the graph. This is justified by the fact that any cycles passing through $u$ also pass through $v$, so that any solution with $u$ in the feedback set can have it exchanged for $v$ without growing the solution. This rule could be implemented in near quadratic time (near linear time for each vertex $u$) using dominator trees\autocite{TarjanDoms}, but this would be a slow check, and initial inspection of the test data suggested it would not often be useful. On the cases where the graph is fully reduced to a cover problem, this rule is equivalent in domination in the cover (which is faster to check), and the majority of the cases where it occurred in practice were already handled by Rule 6 in part 3.3.
    \item Pairwise exclusion - If two vertices $u$ and $v$ are twins, {\em except} that $u$ and $v$ both have edges to the other, then at least one must be excluded (because they form a 2-cycle) and it doesn't matter which (because their other neighbors are identical). Then we can put $u$ in the feedback set and remove it from the graph. This was omitted simply for reasons of development time to run the check for such pairs efficiently, but is expected to improve performance.
\end{enumerate}
After this phase, each subproblem was processed into a cover.

\subsection{Minimal Cycle Rules}\label{sec:mincyc}
It is simple to find {\em some} minimal cycle in a cyclic graph, by running depth-first search to find a cycle and then shortening it along any chords. DFS can in this way enumerate all minimal cycles, but this will typically take exponentially long even on sparse graphs with few minimal cycles, and the problem is expected to be hard in general (see Appendix \ref{app:chordless}). For a strategy to be practical, it should run in polynomial (ideally, near-linear) time, and should reduce the remaining graph afterwards. The rules used in the final solver were:
\begin{enumerate}
    \item Any $C_2$ (a pair of directed edges $(u,v)$ and $(v,u)$) yield a cycle. Any other cycle using one of these edges would have a chord, so both edges can be removed then from the graph. This may lead to collecting extra cycles that have $(u,v)$ as a chord, and that will need to be discarded later, but this rule greatly enhances performance.
    \item Enumerating the chordless cycles of a graph $G$ is equivalent to enumerating chordless cycles of the strongly connected components of $G$. So split $G$ into its SCCs (several separate DFVS constraints) and process each. This also removes any vertices of in- or out-degree zero.
    \item If the graph has a component that is a simple cycle, add that cycle to the set and remove it from the graph.
    \item Check if the graph contains $\{(u,m_1),(m_1,m_2),\dots,(m_k,v),(u,v)\}$, where each $m_i$ has in-degree and out-degree 1. Then any cycle going through an $m_i$ must go through $u$ and $v$ as well, and have a chord. So all the $m_i$ vertices can be removed from the graph.
    \item Check if the graph is has a single vertex $v$ of in-degree two or more. By rule 2, we know that the graph is connected, and that all other vertices have in-degree 1. Then all cycles must pass through $v$. The minimal cycles can be enumerated by depth-first search from $v$, all produced cycles will be chordless, and there will be at most $|V|$ many. The graph can be then be cleared.
    \item The same as rule 5, but with out-degree instead, and all traversing the depth-first search with backedges.
    \item Check if the graph $G$ has a weak articulation point (WAP) $v$, that is, a vertex whose removal will cause the graph to be disconnected, as an undirected graph. If so, let the resulting components be $G_1, G_2, \dots G_k$. Then each chordless cycle in $G$ is contained entirely in $G_i \cup \{v\}$ for some $i$; otherwise it must pass through $v$ twice as it switches from $G_i$ to $G_j$ and back. So, split the DFVS constraint into several components for each $G_i \cup \{v\}$, and enumerate each. Note that these separate constraints share a vertex.
    \item Check if the graph $G$ has an edge $(u,v)$, such that the removal of $u$ and $v$ from the graph would cause it to no longer be connected. If so, let resulting components be $G_1, G_2, \dots G_k$. Then each chordless cycle in $G$ is contained entirely in $G_i \cup \{u,v\}$ for some $i$; otherwise it must pass through one of $u$ or $v$ twice as it switches from $G_i$ to $G_j$ and back, or it passes through each once and has $(u,v)$ as a chord. So, split the DFVS constraint into several components for each $G_i \cup \{u,v\}$, and enumerate each. Note that these separate constraints share a vertex.
    \item If the graph is sufficiently small, try to enumerate minimal cycles by brute force through depth-first search. Once 10 million nodes have been visited in the search tree, the search gives up so that the program may continue.
\end{enumerate}

Rule 7 requires identifying weak articulation points, which can also be found by Tarjan's algorithm. Rule 8 subsumes Rule 7 as a special case, but since it is much slower to check, it is done only after Rules 1-7 fail to apply. Rule 8 is not the same as identifying weak (or strong) bridges in the graph: a bridge leaves the graph disconnected when the edge $(u,v)$ is removed, whereas we consider removing the vertices $u$ and $v$ altogether, including all adjacent edges. This is simply checked by attempting to remove each edge and checking whether the graph stays connected. As a result, Rule 8 is relatively slow, although we suspect there is a linear time algorithm for this check.

The brute force search for Rule 9 was designed to minimize useless exploration and memory allocation. A simple approach is to do depth-first search, returning cycles whenever we encounter a vertex higher in the search tree. Since we only seek chordless cycles, if we arrive at an vertex $v$ that is a child of any vertex $u$ higher in the tree, we can return from $v$ immediately. The check whether $v$ is a child of higher vertices can be done in $O(\textrm{depth})$ or $O(\textrm{indegree}(v))$ time, but this is relatively costly. A bit array of forbidden children can be maintained, but it is unclear when to unmark a vertex as we return up the tree. Our solution was to keep an integer array, counting the number of blocks for each vertex. Visiting a node incremented the blocks on all of its children, returning from a node decremented the blocks, and arriving at a node with more than one block meant we should skip. (Each visited node has one block, from the immediate parent). This algorithm is given in pseudocode in Appendix \ref{app:brute}, and it performed well in practice on the given graphs.

\subsection{Minimum Cover Reduction Rules}
With the graph (usually) reduced to its complete collection of chordless cycles, the problem becomes a minimum cover problem. This is already efficiently amenable to MILP formulation, but further reduction rules play a large role in the performance of the solver. As a cover problem, there is no longer a notion of ordered cycles, but just sets, and we change our word choice here accordingly.

The majority of rules focus on when sets are size 2. When all sets are size 2, this means the original graph consisted entirely of {\em undirected} edges (2-cycles), or at least that with their removal the remaining graph was acyclic. In this case, the problem is minimum vertex cover, which is so well-studied that it has been referred to as the \textit{Drosophila} of parameterized complexity\autocite{Downey_1999,Guo2007}. With a few simple exceptions, our reduction rules are adaptations of those described in Fellows et al.\cite{Fellows2018} and Xiao and Nagamochi\cite{Xiao13}. Several rules refer to ``alternatives", ``folding", ``funnels" and ``desks", which are described in those papers. In the rules below, since we focus on the vertex cover viewpoint, we refer to the ``degree" of a vertex as the number of sets of {\em size two only} that contain it. The vertex may also belong to a ``big set" (of size more than two) or a ``graph" (an unreduced DFVS constraint). If a vertex belongs to no big set or graphs, we call it ``bare".

\begin{enumerate}
    \item If a set $S$ is contained within another set $T$, remove $T$ from the list of sets.
    \item If a vertex $v$ has degree zero and only belongs to one big set (and no graphs), drop it from that big set. Note this may turn a big set into a size two edge.
    \item It a bare vertex $v$ has degree 1, drop $v$ from the problem and adds its neighbor to the cover.
    \item If a bare vertex $v$ has degree 2, check if its neighbors $u$ and $w$ share a set $\{u,w\}$. If they do, $v$ is simplicial and can be dropped, adding $\{u,w\}$. If they don't, and $u$ and $w$ are bare, they can be folded\autocite{Fellows2018}.
    \item If a bare vertex of any degree is simplicial (its neighborhood forms a clique), drop it and add its neighborhood to the set. Note that the neighbors do not need to be bare.
    \item If a vertex $v$ dominates some neighbor $u$, add $v$ to the cover. Here $v$ dominates $u$ if each neighbor of $u$ is also a neighbor of $v$, each big set of $u$ is also a big set of $v$, and $u$ does not belong to a graph.
    \item Check for a $k$ funnel $u-v-Nv'$, where $u$ and the set $Nv'$ comprise the neighborhood of $v$ and $Nv'$ is a clique. If $v$ is bare and $u$ doesn't belong to a graph, we can resolve the funnel as an alternative. As part of the alternative, for each big set $C$ that $u$ belongs to and each vertex $w$ in $N(v)\setminus N(u)\cup\{u\}$, we remove $C$ from our list of sets, and add $C\cup\{w\}\setminus\{u\}$. This replication is analogous to the new edges added for an alternative.
    \item Each vertex is checked for confinement, as per Xiao and Nagamochi\cite{Xiao13}. If $v$ is unconfined (looking only at the edges), and the implicated set $S$ and its neighborhood $N(S)$ are entirely bare, then $v$ is truly unconfined and we include it in the cover.
    \item Check for a desk $(a,b,c,d)$, also as per \cite{Xiao13}. If the desk and all its neighbors are bare, they form an alternative that can be folded in.
\end{enumerate}

When a vertex belonged to a graph, it would never be dropped (safely excluded from the cover), but it would sometimes be included in the cover and removed from the graph. This triggered further some kernelization of the graph, removing in- or out-degree zero vertices from the graph.

There are several improvements to be made. Rules 7 likely requires less bareness to apply, as does Rule 8, which can use the same big set replication trick as Rule 6. A mistake in applying Rule 6 led to the bug that invalidated our initial submission: can we handle alternatives one of them (e.g. $u$ in Rule 6) belongs to a graph $G$? As we replicate the big sets that $u$ belongs to with its alternatives $N(v)\setminus N(u)\cup\{u\}$, we could replicate $u$ in $G$ by adding in-edges and out-edges for each $N(v)\setminus N(u)\cup\{u\}$. This is valid when $N(v)\setminus N(u)\cup\{u\}$ is disjoint from $G$, but when it intersects $G$, this can lead to incorrect cycles. Detecting and handling the case where $N(v)\setminus N(u)\cup\{u\}$ is disjoint from $G$ is another avenue for improvement. Finally, when graphs were updated due to vertex inclusion, rerunning the graph rules (e.g. SCC splitting) and cycle rules (hoping to find minimal cycles) could be useful. There were additional reduction rules we considered but did not apply:

\begin{enumerate}
    \item Crown Reductions - These are well studied within vertex cover, e.g. Chlebik et al.\cite{Chlebik08}, but they would need to be carefully adapted for non-bare vertices, and are complicated to detect efficiently.
    \item Twins - If $k$ independent vertices of degree $k+1$ all have identical neighbors, they form a simpler case of a crown reduction. We implemented a check for these but did not encounter them in a single test instance and so left the check disabled.
    \item Generalized Desks - A desk $(a,b,c,d)$ is an induced a 4-cycle of vertices degree at least three and at most four, with the property that $|N(a)\cup N(c)| \le 4$ and $|N(b)\cup N(d)| \le 4$, in which case $\{a,c\}$ and $\{b,d\}$ form an alternative. Upon consideration, this definition is unnecessarily restrictive. It is sufficient to say that $(a,b,c,d)$ induces a 4-cycle (of any degree vertices), where $|N(a)\cup N(c)| \le 4$, $|N(b)\setminus N(d)| \le 1$, and $|N(d)\setminus N(b)| \le 1$. For a proof, see Appendix \ref{app:gendesk}. We implemented a check for these generalized desks, but did not encounter them in test instances and left them disabled in the final submission.
    \item Co-matching into a clique - Reduction Rule R.8 in Fellows et al.\cite{Fellows2018}. This needs to be carefully adapted for non-bare vertices, and even the author comment on "its complexity".
    \item Low degree rules - Reduction Rules R.9 through R.12 in Fellows et al.\cite{Fellows2018}. These also need to be carefully adapted for non-bare vertices.
\end{enumerate}

\subsection{Solving the MILP}
When the resulting problem has no DFVS constraints, there is essentially only one way to provide the problem to the MILP solver as a linear program with binary variables. Plugins in SCIP handle automatic interconversion between linear constraints, packing constraints, detecting odd cycles and cliques, and so on. The more challenging case is when there are still parts of the graph remaining.

With the problem not yet in the form of a linear program, the DFVS constraints must be lazily enforced. Significant effort went into finding the best way to interact with SCIP in the most productive and cooperative way, including lazy constraint enforcement at nodes in the search tree, or providing heuristic solutions as upper bounds. Ultimately these were unproductive, because the very intelligent "presolver" in SCIP did not have complete information on the problem: when it was informed that there were lazy constraints not expressed by the MILP, it could not do many of its most useful tricks. The alternative constraint enforcement loop of "solve the problem to optimum, find new cycles, and repeat" turned out to be the most efficient. When SCIP solved the relaxed MIP and found a candidate optimum, the resulting graph (with the candidate feedback set removed) was checked for acyclicity. If it had cycles, a maximal set of edge-disjoint cycles were added to the MILP and it was resolved. Otherwise, a true optimum feedback set had been found.

Significant improvement was found when an initial random search through the graph to just find {\em some} large list of cycles as a starting point. Since often two or three repetitions was needed, a random list of cycles could often shorten it to only require one or two repetitions. On the other hand, too many initial cycles lead to performance degradation, presumably due to SCIP processing the very large number of constraints; for the graphs present in the competition, finding 50 random cycles at each step proved useful.

Generated cycles were found through a randomized DFS exploration. The found cycle would then be trimmed at any chords until it was chordless Finally, a random edge from the cycle would be removed from (the temporary copy of) the graph to prevent it being found again in this round, and the new a new cycle was sought. This continued until the temporary copy of the graph was acyclic. This would then be executed 50 times with different random seeds. This happened once before the first MILP solving, and once after each candidate optimum was found.

\section{Conclusion}
We built a high-performing exact solver for Directed Feedback Vertex Set without needing to implementing and branch-and-bound or search heuristics. We optimistically consider this evidence that MILP solvers are so powerful and flexible, that the best approach to the problem is to translate into MILP, and focus one's effort on improving and refining that translation as much as possible, rather than developing a search framework from scratch. We also observe that the work invested in Vertex Cover has excellent applicability to other problems, in the cases that they can be phrased as cover problems; again, effort may be best spent on reducing to well-studied problems, rather than developing custom reduction rules or search strategies for the unique problem.

\printbibliography

\appendix

\section{Hardness of Chordless Cycles}\label{app:chordless}
Enumerating all chordless cycles in a directed graph is a tall order, as there can be exponentially many, but we could hope for an algorithm that runs in $O(poly(|V|)*k)$ time, where $k$ is the number of chordless cycles to find, i.e. output-polynomial time. This is quite unlikely, as even checking whether some chordless cycle passes through a given cycle is NP-complete, as Anna Lubiw proved in 1988\autocite{Lubiw88}. This doesn't in itself preclude the possibility of an output-polynomial time algorithm, even under $\mathsf{P}\neq\mathsf{NP}$ as the relevant graphs have exponentially many chordless cycles; however, it does pose a substantial obstruction to any such algorithm, essentially forbidding search-based backtracking algorithms. Related problems such as detecting odd chordless cycles, even chordless cycles, or chordless paths through three vertices were also shown to be NP-complete. For undirected graphs, output-polynomial time algorithms are known\autocite{Read75,Ferre14,Uno14}, which could be useful in the {\em undirected} Feedback Vertex Set problem.

\section{Brute Force Cycle Enumeration}\label{app:brute}
This appendix contains the pseudocode for Rule 9 in \ref{sec:mincyc}. The Java implementation can be found \href{https://github.com/Timeroot/DVFS_PACE2022/blob/4458649a59bdff168c104d04c9f677816ddfbf7a/src/ChordlessCycleEnum.java#L13}{in this GitHub revision}.
\begin{algorithm}[H]
\caption{Depth-First Chordless Cycle Enumeration}
\begin{algorithmic}
\Procedure{EnumChordlessCycles}{$G$}
\State $\textsf{NodeCount} \gets 0$
\State $\textsf{Path} \gets []$
\For{$i \gets 1$ to $V$}
    \State $\textsf{Blocks}[i] \gets 0$
\EndFor
\For{$\textsf{Start} \gets 1$ to $V$}
    \State $\textsf{Blocks}[\textsf{Start}] \gets -3N$\Comment{Start will never be blocked}
    \State \Call{CycleHelper}{$G$,\textsf{NodeCount},\textsf{Path},\textsf{Blocks},\textsf{Start}} \Comment{Call recursive subroutine}
    \State $\textsf{Blocks}[\textsf{Start}] \gets 1$\Comment{Subsequent subroutine calls should omit this vertex}
    \If{$\textsf{NodeCount}>10^7$}
        \State \Return false \Comment{Indicating incomplete enumeration}
    \EndIf
\EndFor
\State \Return true\Comment{Complete enumeration}
\EndProcedure
\Procedure{CycleHelper}{$G$,\textsf{NodeCount},\textsf{Path},\textsf{Blocks},$v$}
\State $\textsf{NodeCount} \gets \textsf{NodeCount}+1$
\If{$\textsf{NodeCount}>10^7$} \Return
\EndIf
\If{$v = \textsf{Path}[1]$}\Comment{Cycle found}
    \State \textbf{output} \textsf{Path}
    \State \Return
\EndIf
\If{$\textsc{Length}(\textsf{Path}) > 0$}\Comment{Any time except the first call}
    \State $\textsf{Blocks}[v] \gets \textsf{Blocks}[v]+1$
    \For{$p \gets \textsc{BackEdges}(v)$}
        \State $\textsf{Blocks}[p] \gets \textsf{Blocks}[p]+1$
    \EndFor
\EndIf
\For{$c \gets \textsc{Edges}(v)$}
    \State $\textsf{Blocks}[c] \gets \textsf{Blocks}[c]+1$
\EndFor
\State $\textsc{Append}(\textsf{Path},v)$

\If{$\textsc{Contains}(\textsc{Edges}(v),\textsf{Path}[1])$}\Comment{If start is a child, it is our only option}
    \State \Call{CycleHelper}{$G$,\textsf{NodeCount},\textsf{Path},\textsf{Blocks},$\textsf{Path}[1]$}
\Else
    \For{$c \gets \textsc{Edges}(v)$}
        \If{$\textsf{Blocks}[c] \le 1$}
            \State \Call{CycleHelper}{$G$,\textsf{NodeCount},\textsf{Path},\textsf{Blocks},$c$}
            \If{$\textsf{NodeCount}>10^7$} \Return
            \EndIf
        \EndIf
    \EndFor
\EndIf
\State $\textsc{RemoveLast}(\textsf{Path})$
\If{$\textsc{Length}(\textsf{Path}) > 0$}\Comment{Any time except the first call}
    \State $\textsf{Blocks}[v] \gets \textsf{Blocks}[v]-1$
    \For{$p \gets \textsc{BackEdges}(v)$}
        \State $\textsf{Blocks}[p] \gets \textsf{Blocks}[p]-1$
    \EndFor
\EndIf
\For{$c \gets \textsc{Edges}(v)$}
    \State $\textsf{Blocks}[c] \gets \textsf{Blocks}[c]-1$
\EndFor

\EndProcedure
\end{algorithmic}
\end{algorithm}

\section{Generalized Desks}\label{app:gendesk}
The Desk reduction rule uses the fact that $\{a,c\}$ and $\{b,d\}$ form alternative sets\autocite{Xiao13} in vertex cover, in other words, there exists an optimal solution which contains all of one set and none of the other. They are said to form a desk if each vertex is degree 3 or 4, both sets have at most 4 neighbors, and the four vertices induce a 4 cycle $a-b-c-d-a$.

We see that a desk is a valid alternative as follows (following the proof of Xiao and Nagamochi\cite{Xiao13}). First, at least one of the two sets must be in contained in any vertex cover, otherwise (wlog) $c$ and $d$ are not in the cover, and the edge $(c,d)$ isn't covered. If both sets are contained in the vertex cover $S$, then the set the $S' = S\cup N(\{a,c\})\setminus \{a,c\}$ is also a cover. Since $N(\{a,c\})$ is at most size 4, and it already includes $b$ and $d$, $|S'| \le |S|$. If a cover $S$ contains $\{a,c\}$ and an extra vertex $b$, then we use $S' = S\cup N(b)\setminus \{b\}$. Either $N(b)$ is size 3, has one vertex besides $a$ and $c$, and $|S'| \le |S|$; or $N(b)$ has size 4, $b$ must have three neighbors in common with $d$ (since $d$ has degree at least three), and $|S'| \le |S|$. Since any cover that doesn't obey the alternative property can be transformed into a new cover that does and at least as small, this is a valid alternative.

The notion of desks can then be somewhat generalized, because we did not use all the conditions. We need $|N(a)\cup N(c)| \le 4$ for {\em one} of the two pairs, to handle the case where $|S \cap \{a,b,c,d\}| = 4$. And we need that $b$ and $d$ each have at most one neighbor that the other does not (and likewise for $a$ and $c$), to handle the case where $|S \cap \{a,b,c,d\}| = 3$. If we fix $\{a,c\}$ to be the pair with $|N(a)\cup N(c)| \le 4$, then $b$ and $d$ can be of arbitrarily high degree, as long as $|N(b) \setminus N(d)| \le 1$ and $|N(d) \setminus N(b)| \le 1$.

\begin{definition}
An induced 4-cycle $(a,b,c,d)$ is a {\em generalized desk} if:
\begin{itemize}
    \item Each vertex is degree 3 or greater
    \item $|N(a)\cup N(c)| \le 4$
    \item $|N(b) \setminus N(d)| \le 1$
    \item $|N(d) \setminus N(b)| \le 1$
\end{itemize}
\end{definition}

Note that if all dominating vertices have already been removed from the graph, then the $< 1$ conditions cannot occur in the last two points. And as we've showed,

\begin{proposition}
If $(a,b,c,d)$ forms a generalized desk, then $\{a,c\}$ and $\{b,d\}$ are alternative sets. A vertex cover problem can be reduced by removing all four vertices from the graph, adding edges $(u,v)$ for each $u\in N(\{a,c\})$ and $v\in N(\{a,c\})$, and increasing the size of the cover by two (i.e. reducing the parameter $k$ by two).
\end{proposition}

\end{document}